\documentclass[prb,11pt]{revtex4-1}

\usepackage{amsmath}    
\usepackage{graphicx}   
\usepackage{verbatim}   
\usepackage{color}      
\usepackage{subfigure}  
\usepackage{hyperref}   
\newcommand{\nl}{\\ \nonumber &&}
\begin{document}

\title{Removing Ostrogradski ghost from degenerate gravity theories}
\author{Biswajit Paul }

\address{ National Institute of Technology Agartala, India }

\address{\tt 
	biswajit.thep@gmail.com}

\begin{abstract}

The Ostrogradski ghost problem that appears in higher derivative system is considered for theories with constraints. A new prescription for removal of the ghost creating momenta that { come along }the constrained systems is described based on the Dirac's constraint analysis. It is shown how one can make the canonical Hamiltonian bounded from below by systematically removing the constraints appearing in the system  thereby reducing the effective dimension of the phase space. To show the effect of higher derivative terms we consider  the singularity free Gauss-Bonnet theory coupled via a matter field to the Einstein Hilbert action. Finally we construct the canonical Hamiltonian for the theory that is bounded from below.
\end{abstract}
\maketitle 

 \section{Introduction}
  Higher derivative(HD) theories have  persuaded the physics community for their usefulness  in the field theoretic context. By higher derivative we  mean that the fields appearing in the action have time derivative more than one. Historically these HD terms were added to the Lagrangian for renormalisation \cite{stelle,muneyuki, buch}, perturbative corrections\cite{pert}. Also they frequently appear in diverse fields like relativistic particle \cite{nesterenko}, string theory \cite{string} and general relativity \cite{myers, codello, mignemi, camps}.

    Right from the inception of the higher derivative theories they were diagnosed with a problem called the Ostrogradski ghost problem \cite{ostro, woodard1}. The ghost fields are nothing but some unphysical field arising in the theory that give rise to the  negative norm states while quantising the theory. More precisely, in the canonical Hamiltonian, there may appear terms linear in momenta  of the higher derivative fields. Due to this, the Hamiltonian is not bounded from below which exactly creates the problem as the negative energy states propagate the whole phase space and give rise to instabilities. Although Ostrogradski pointed out that only non-degenerate theories will have the ghost problem but here we show that, at the classical level, degenerate theories will also consist of terms in the canonical Hamiltonian that cause the instabilities. So, in the case of  both  degenerate and non-degenerate theories the Ostrogradski ghost problem can appear at the classical level. These instabilities some times are of tachyonic nature depending upon the wave function, if it is oscillatory. 
  
    Over  many decades there were numerous attempts to solve the problem of these negative norm states by different authors. Like, one possible way was suggested for removing the ghost fields  by applying some boundary conditions e.g.  in \cite{maldacena} the ghost free version was obtained by applying the Neumann boundary condition in the wave function.  Also, Bender and Manheim in \cite{bender} showed that for a specific class of theories with $PT-$symmetry the ghost fields behave as usual fields and give positive $PT-$norms. For the theories with no constraints, it was suggested in \cite{chen} to consider some external relations between the phase space variables, thereby decreasing the number of degrees of freedom. Recently, the authors in \cite{hassan} has shown that for massive and bimetric gravity theories there appear two second class constraints which in turn help to elliminate the ghost field and its corresponding momenta. There are many work done very recently to find ghost-free massive gravity theories \cite{massive_gf}. Not always the ghost fields are `Bad', in some cases, despite they are present in the theory they do not pollute it. As in \cite{volkov}, the massive gravity theory was having a negative energy state but it was from a disjoint branch and hence can not communnicate to the positive energy states thus leading to no instability. There are also attempts to remove the ghosts by considering  an infinite set of higher derivative terms, in particular in the form of an exponential\cite{majumdar}. In \cite{cho}, the Ghost free states were found for the linearised gravity with the Gauss-Bonnet couplings in the Randal-Sundurum picture. For degenerate gravity theories Ostrogradski ghost removal was discussed in \cite{langlois}.

	   The Einstein-Hilbert(EH) action by definition contains higher derivative terms of the field $g_{\mu\nu}$  but it is easy to point out that these higher derivative terms actually are surface terms  that can be neglected while considering the integrations. On the other hand,  usual gravity theories are not renormalisable unless higher derivative terms are added \cite{stelle}. For that reason the higher derivative terms are inevitable in gravity theories.  There are two ways by which HD terms enter into the gravitational action. One directly as function of higher curvatures  like $f(R)$ or via the matter fields. These type of models of adding matter fields is a direct implication that one can get from the compactification of Kaluza-Klein theories. In this paper we will restrict ourselves only in the higher curvatures.  For that, as a viable HD term, we consider the Gauss-Bonnet gravity. The Gauss-Bonnet term is a special combination of the  curvature squared terms ($R^2, R^{\mu \nu}R_{\mu \nu}, R^{\mu \nu \rho \sigma}R_{\mu \nu \rho \sigma}$) which is actually a surface term added to the action. But it is very important for explaining dark energy \cite{amendola}, inflationary scenario \cite{lidsey}, bouncing universe \cite{maeda},   implications of massive gravity \cite{hendi}, brane world gravity \cite{lidsey1} and many more\cite{GB}. Very recently the coupling parameters for the Gauss-Bonnet terms were  proposed based on the  results of BICEP2 and Planck \cite{nupane}. Discussions on the stability conditions of the vacuua in Gauss-Bonnet gravity can be found in \cite{charmousis}. 
   
    In this paper, we first describe, in detail, how the ghosts appear via the canonical Hamiltonian and thereafter describe a procedure for removing these ghosts. To apply this method in models we first consider Einstein-Hilbert action and then add  Gauss-Bonnet term which is coupled   via a field dependent coupling parameter \cite{kanti, rizos}. It is worth to be mentioned that although these two models are independent of the ghosts but at first glance the ghost creating momenta appear in the canonical Hamiltonian and make the theory to appear as if  ghost dependent. The method described in this paper is well capable of removing these ghosts from the system.  The Gauss-Bonnet model have singularity free solution with conditions on the coupling parameter. With the FRLW background we found the metric in the preferred minisuperspace version after ADM decomposition \cite{ADM,  cordero1,BMP_rt}. Being a higher derivative system, we adopt the  first order formalism \cite{BMP, deriglazov} and rename all the field variables to apply the Hamiltonian formalism. We found out that the system has constraints and therefore  followed the Dirac constraint analysis to find out all the constraints in the system \cite{dirac,sunder}. We construct the canonical Hamiltonian and find that there are terms   linear in momenta of the fields. The momenta which corresponds to the higher derivative fields can give rise to negative norm states. To get rid of them, we elliminate these momenta  from the canonical Hamiltonian. Accordingly, we remove the momenta appearing in the constraints by solving the second class constraints. It is to be noted that the ghost creating momenta will be appearing in some of the constraints which may be first class second class in nature. If they are second class, we  can solve the constraints for the momenta and replace it in the canonical Hamiltonian. On the other hand if the momenta appear in the first class constraints we need to introduce gauge conditions. For this model, all the constraints obtained are found to be  second class in nature.  Essentially, after solving the constraints all the Poission brackets in the system has to be replaced by Dirac brackets which, during quantisation, will play the role of commutators.  
    Based on this concept, for theories with constraints, we showed how to construct systematically the ghost free Hamiltonian  that is independent of linear dependence of the momenta corresponding to the  higher derivative field(s). {The prescription presented here  may be useful in models from different fields although a general proof is at warrant to value the method more. }

   The paper is organised as described. In Sec II we describe  the general procedure for how the momenta appear by default in the canonical Hamiltonian for a higher derivative theory. We also describe, for a constraint theory, how to remove these fields by considering the second class constraints in the reduced phase space. In Sec III we consider the  the Einstein Hilbert action with the FRLW background spacetime. Adopting  the first order formalism we perform the Hamiltonian formulation and finally found out the ghost free version of the canonical Hamiltonian. In sec IV we add the Gauss Bonnet term  via a matter field to the Einstein Hilbert action and construct the canonical Hamiltonian in the first order formalism. We solve the second class constraints appearing in the system  to found out the final form of the ghost free canonical Hamiltonian. Also we compute the corresponding Dirac brackets between the canonical variables in the reduced phase space.
             
 \section{Open ended higher derivative theories: general prescription}
 Lagrangian of a higher derivative(HD) theory of n-th order derivative in time  is  generally written as
 \begin{equation}
 L=f(q, \dot{q}, \ddot{q}, .... q^{(n)}).
 \end{equation}
 The Lagrangian is a function in the configuration space which consist of the field (q) and its  time derivative(s).  We convert the Lagrangian into first order theory by incorporating the variables 
 \begin{equation}
 q_1=q, q_2 = \dot{q}\ \ \  .... \ \ \ \  \ \  \ q_{n} = q^{(n-1)}.
 \end{equation}
 So, the Lagrangian in the first order is written as 
 \begin{equation}
 L^\prime = F(q_1, q_2, q_3..... \dot{q}_{n}) + \sum_{i=1}^{ n} \lambda_i (q_{i+1} -\dot{q}_i ).
 \end{equation} 
Here these $\lambda_i$ are the Lagrange multipliers multiplied to the constraints that appeared dute to the redefinition. Consequently, the dimension of the configuration space is increased.  The momenta in this formalism can be written as
 \begin{eqnarray}
  p_i &=& \frac{\partial L^\prime}{\partial \dot{q}_i} = \frac{\partial F}{\partial \dot{q}_i} - \lambda_i, \ \ \ \ \ \  i=1,2 \dots (n-1) \label{pri1} \\    
 p_{\lambda_i} &=& \frac{\partial L^\prime}{\partial \dot{\lambda}_i} = 0. \label{pri2}
 \end{eqnarray}
 The constraint $p_{\lambda_i}$ is obvious and will be generated every time while defining the new fields. However, we should not bother about this as these fields are nondynamical and can eventually be set up to zero at the end. The character of $p_i$'s is important. The function $\frac{\partial F}{\partial \dot{q}_i}$ is zero other than for $i=n$  depending on the nature of $F(q_1, q_2...\dot{q}_n)$.
 The poission brackets between the canonical variables $ Q_i \equiv \{ q_i, q_{\lambda_i}\}$ and their corresponding momenta $ P_i \equiv \{ P_i, P_{\lambda_i}\}$ are
 \begin{equation}
 \{q_i ,p_j\} = \delta_{ij}. \label{pb_def}
 \end{equation} Surely (\ref{pri1},\ref{pri2})  leads to primary constraints and hence the momenta will not be a function of the derivative of the corresponding field.  Thus the primary constraints are written generally as
 \begin{equation}
\Phi_{i1} : p_{\lambda_i} \approx 0, \Phi_{i2} : p_i - \frac{\partial F}{\partial \dot{q}_i} + \lambda_i \approx 0. \label{pricons}
 \end{equation}
 
 The canonical Hamiltonian is
 \begin{eqnarray}
 	\nonumber
 	 H_c &=& p_1 \dot{q}_1 + p_2 \dot{q}_2 + p_\lambda \dot{q}_\lambda - L^\prime. \nonumber \\
 	 &=& \sum_{i=1}^{n-1}p_i q_{i+1} + h(q_1, q_2 ... q_n) \label{canham1}
 	\end{eqnarray}
 In  the canonical Hamiltonian  (\ref{canham1}) the first term consist of (p,q) but rest of the terms will also be simplified to make the Hamiltonian a function of the phase space variable once we consider the explicit form of the functions $h(q_1, q_2 ... q_n)$.  What is important to look in the Hamiltonian is that there are terms proportion to $p_i$. These terms can span in the whole phase-space region and consequently make the canonical Hamiltonian negative. This states that the Hamiltonian is not bounded from below and can range between both the positive  or negative axis i.e. with an open ended solution. Therefore, in the quantum picture of the theory there appear negative norm states which are known as ghost states. This is essentially a manifestation of the Ostrogradski theorem. It is interesting to see that renaming the variable to look it as if a first order theory and also by defining the momenta in the usual way did not help us to get rid of the negative Hamiltonian. This general discussion reveal that appearance of the HD momenta to make the canonical Hamiltonian negative is inherent character for any HD theory.
 
 \subsection{Removing ghost degrees of freedom using constraints}
 
 What we understood till now is that  in the HD theory, there will eventually appear the Ostrogradski ghosts and that is evident from the canonical Hamiltonian. As mentioned in the introduction, for their removal, there are several methods available in the literature for the non-degenerate theories. But for degenerate theories a less introspected way may be followed when the ghost creating momenta are involved in the constraints.  In (\ref{pricons})  we get the set of primary constraints where the set of  momenta $p_i$ appear. Also we notice that these set of momenta $p_i$ appearing in the canonical Hamiltonian are responsible for  creating the negative norm states. A way out can be removal of these momenta by solving the constraints as at the end in the physical phasespace the constraints will eventually be removed. For that it is essential that we should get the full constraint structure.   Let us write down a general form for the total Hamiltonian which is given as 
 \begin{equation}
 H_T = H_c + \sum^{n}_{i=1}u_i \Phi_{i1} + \sum^{n-1}_{i=1}u_i \Phi_{i2}
 \end{equation}
 The evolution of the primary constraintsmay give rise to secondary constraints and in the similar way we get tertiary constraints and so on. Once we have all the constraints we can further categories them as first class or second class. This division is essential as we can know about the gauge symmetries of the system also \cite{dirac, BMP}. Solving the constraints let us reduction of  dimension of the phase space, a way out  mentioned in \cite{chen} for non-degenerate HD theories. The same can be done by the expense of the second class constraints. Now there may arise two conditions
 \begin{itemize}
 	\item \textit{The Canonical Hamiltonian is a first class constraint.} This means there will be  no more generation of the constraint. Since the theory has first class constraints at the primary level there is gauge degree of freedom. The dimension of the phase space can be reduced  by incorporating external condition in ad hoc basis. These are called gauge conditions. These gauge conditions make the first class constraints second class. Next we solve all these second class constraints and incorporate Dirac brackets. The canonical hamiltonian becomes free from the momenta which were the source of instabilities.  
 	
 	\item \textit{The Canonical Hamiltonian is a second class constraint.} In this case there will be more constraint arising in the theory. The reduced dimensionality of the phase space can be obtained by setting these second class constraint to zero. The poission brackets of the theory to be replaced by Dirac brackets.
 	
 \end{itemize} 
 
So reduction of the degrees of freedom is a must. In the squeezed phasespace proper choice of the canonical variables can be done by inspecting the corresponding brackets. There after it will be  a ghost free  theory.  Between any two functions $f$ and $g$ of the canonical variables  Dirac brackets are defined as 
\begin{equation}
\{f,g\}_D=\{f,g\} - \{f,\Psi_i\}\Delta^{-1}_{ij}\{\Psi_j , g\}. \label{DB_def}
\end{equation}
Where $\Delta_{ij}$ is the Poission bracket matrix of the second class constraints. 
 
 \section{ Einstein Hilbert action in minisuperspace}
 We consider the following metric of FRLW kind as   
 \begin{equation}
  ds^{2} = - dt^{2} + da^{2} + a^2 d\Omega_{3}^{2},
 \end{equation}  
 where $d\Omega_{3}^{2}$ is the metric for unit 3 sphere. We parametrise the brane using the parameter $\tau$ as 
 \begin{eqnarray}
 x^{\mu} = X^{\mu}(\xi^{a}) = \left( t(\tau),a(\tau), \chi, \theta, \phi \right),
 \end{eqnarray}
 $a(\tau)$ is known as the scale factor.
 
 After ADM decomposition  with space like unit normals ($N(\tau)=\sqrt{\dot{t}^{2}-\dot{a}^{2}}$ is the lapse function)
 \begin{equation}
 n_{\mu} = \frac{1}{N}(-\dot{a}, \dot{t}, 0,  0,0),
 \end{equation}
 the induced metric on the world volume is given by,
 \begin{equation}
 ds^{2} = -N^{2} d\tau^{2} + a^{2} d \Omega_{3}^{2}. \label{metric}
 \end{equation}
 Computation of the Ricci scalar is straight forward  which is given by
 \begin{equation}
 \mathcal{R} = \frac{6 \dot{t}}{a^{2} N^{4}}(a \ddot{a}\dot{t} - a\dot{a}\ddot{t} + N^{2}\dot{t}).
 \end{equation}
The Lagrangian corresponding to the standard Einstein-Hilbert action with the non-zero cosmological constant is 
 \begin{equation}
 \mathcal{L} = \sqrt{-g} \left(\frac{\alpha}{2}\mathcal{R} - \Lambda\right).
 \end{equation} 
 The Lagrangian in terms of of the fields with  arbitrary parameter $\tau$ can be written as\cite{cordero1, BMP_rt}
 \begin{equation}
 L(N,N',a, a', a'' ) = \frac{ a}{ N(\tau)^2}  \left(-3 a a' N'(\tau )+3 N(\tau) \left(a a''+a'^2\right)+N(\tau)^3 \left(\Lambda  a^2+3\right)\right).
 \label{hdlag}
 \end{equation}
  
Here `prime' refers to differentiation with respect to parameter $\tau$. Note that the Lagrangian (\ref{hdlag}) contains higher derivative terms of the field $a$. However we can write it as \cite{cordero1}
 \begin{equation}
 L= -\frac{a{{a'}}^2}{N} + aN\left(1 - a^2H^2\right) + \frac{d}{d\tau}\left(\frac{a^2{a'}}{N}\right).
 \label{orglag}
  \end{equation}
  The above lagrangian (\ref{orglag}) has total derivative term which acutually vanishes while performing the integrations, but we will keep this term as it carries information about the entropy of the system. With the following redefinition of the fields \begin{equation}
  a'(\tau)=A(\tau) \label{redefinition}
  \end{equation}
  we obtain the first order Lagrangian which is given by 
  \begin{eqnarray}
  \mathcal{L}=\frac{a}{N(\tau)^2} \left(3 N(\tau) \left(a A'+A^2\right)-3 a A N'(\tau )+N(\tau)^3 \left(\Lambda  a^2+3\right)\right)+ \lambda _a \left(A-a'\right).
  \end{eqnarray}
 Here we incorporated the constraint due to field redefinition via the Lagrange multiplier $\lambda_a(\tau)$.    The Euler -Lagrange equation of motions  are
 \begin{eqnarray}
a(\tau ): && \frac{6 a A'}{N(\tau )}-\frac{6 a A N'(\tau )}{N(\tau )^2}+\lambda _a'+3 N(\tau ) \left(\Lambda  a^2+1\right)+\frac{3 A^2}{N(\tau )}=0 \\
A(\tau ):  && \frac{6 a \left(A-a'\right)+N(\tau ) \lambda _a}{N(\tau )}=0 \\ 
N(\tau ): && \frac{a \left(6 A a'+N(\tau )^2 \left(\Lambda  a^2+3\right)-3 A^2\right)}{N(\tau )^2}=0 \\ 
\lambda _a(\tau ): && A-a'=0
 \end{eqnarray}
 We construct the phasespace in the next subsection for Hamiltonian formulation. 
 \subsection{Constructing ghost free Hamiltonian}
  The  phase space is constructed out of the variables  $\{ N(\tau), a(\tau), A(\tau), \lambda_a(\tau) \}$ and their corresponding momenta   are $\{ \Pi_N(\tau), \Pi_a(\tau), \Pi_A(\tau), \Pi_{\lambda_a}(\tau) \}$. Since the Lagrangian is in first order form, the momenta defined in the usual way are
 \begin{eqnarray}
 \nonumber &&
 \Pi _{N}=-\frac{3 a^2 A}{N(\tau )^2}, \nl 
 \Pi _{a}=-\lambda _a , \nl 
 \Pi _{A}=\frac{3 a^2}{N(\tau )}, \nl 
 \Pi _{\lambda _a}=0.
 \end{eqnarray}
 None of the momenta here are invertible with respect to the corresponding velocity, hence we can construct the primary constraints as 
 \begin{eqnarray}
 \nonumber && 
 \Phi _1= \Pi _N+\frac{3 a^2 A}{N(\tau )^2} \approx 0, \nl 
 \Phi _2=\Pi _a+\lambda _a \approx 0, \nl 
 \Phi _3=\Pi _A -\frac{3 a^2}{N(\tau )} \approx 0, \nl 
 \Phi _4=\Pi _{\lambda _a}  \approx 0.
 \end{eqnarray}
 The nonzero Poission brackets between the primary constraints are 
 \begin{eqnarray}
 \nonumber &&
 \{\Phi_1,\Phi_2 \}=\frac{6 a A}{N(\tau )^2}, \nl 
 \{\Phi_2,\Phi_3\}=\frac{6 a}{N(\tau )}, \nl
 \{\Phi_2,\Phi_4\}=1.
 \end{eqnarray}
It seems that all the primary constraints are second class in nature. If we replace $\Phi _1$ by the combination $ \xi_1 = \Phi _1 +\frac{ 6 a A}{N(\tau)^2}\Phi _4$ we get $\xi_1$ as first class at this level. But we are not interested in exploring the gauge symmetries, so let us make use of these second class constraints. 
 Next we construct the canonical Hamiltonian which is given by
 \begin{equation}
 H_{can} = -\frac{3 a A^2}{N(\tau )}+A \Pi _a -a N(\tau) \left(\Lambda  a^2+3\right). \label{canham2}
 \end{equation}
Surely, this Hamiltonian is not bounded from below due to the existence of the momenta $\Pi_a$. Now the total Hamiltonian is given by 
\begin{equation}
H_T = H_{can}+ \Lambda_1 \xi_1 + \Lambda_2 \Phi_2 + \Lambda_3 \Phi_3 + \Lambda_4 \Phi_4
\end{equation}
To see the time evolution of the primary constraints we compute the Poission brckets between the primary constraits and total Hamiltonian which give
\begin{eqnarray}
 &&
\{\xi_1, H_T \} = a \left(\Lambda  a^2+\frac{3 A^2}{N^2}+3\right),\label{pcons_time1} \\ &&
\{\Phi_2, H_T \} = \frac{3}{N(\tau )}  \left(2 \Lambda _3 a + N(\tau )^2 \left(\Lambda  a^2+1\right)+A^2\right){N(\tau )}+\Lambda _4 \label{phi2_HT} \label{pcons_time2} \\ && 
\{\Phi_3, H_T \} = - \frac{6 a \left(\Lambda _2-A \right)}{N(\tau )} + \lambda _a \label{pcons_time3} \\ &&
\{\Phi_4, H_T\} = -\Lambda _2 +A_2 \label{pcons_time4}
\end{eqnarray}
Equating (\ref{pcons_time4})them to zero we get $\Lambda_2 = A_2$. Now demanding (\ref{pcons_time3}) and (\ref{pcons_time1}) to zero and using $\Lambda_2=A_2$ we get two secondary constraints
\begin{eqnarray}
 &&
\Psi _1=a \left(\Lambda  a^2+\frac{3 A^2}{N(\tau)^2}+3 \right)\approx 0, \label{eh_scons1} \\ &&
\Psi _2 = \lambda _a \approx 0. \label{eh_scons2}
\end{eqnarray}
 Non-zero Poission brackets between the primary and secondary constraints are 
 \begin{eqnarray}
 \nonumber &&
 \{\Phi_1,\Psi_1\}=-\frac{6 a A^2}{N(\tau )^3} ,\nl 
 \{\Phi_1,\Psi_2\}=-\frac{6 a A }{N(\tau )^2} \nl 
  \{\Phi_2,\Psi_1\} = 3 \Lambda  a^2+\frac{3 A^2}{N(\tau )^2}+3, \nl
 \{\Phi_3,\Psi_1\} =\frac{6 a A}{N(\tau )^2}, \nl 
 \{\Phi_4,\Psi_1\} = -1. 
 \end{eqnarray}
 Time conservation of the secondary constraints (\ref{eh_scons1}) and (\ref{eh_scons2}) give
 \begin{eqnarray}
 \nonumber &&
 \{\Psi_1, H_T\} = \frac{6 \Lambda _1 a A^2}{N(\tau )^3}-\frac{6 \Lambda _3 a A }{N(\tau )^2}+\Lambda _2 \left(-3 \Lambda  a^2-\frac{3 A^2}{N(\tau )^2}-3\right) ,\\ && 
 \{\Psi_2, H_T\} = \frac{6 \Lambda _1 a A}{N(\tau )^2}+\Lambda _4. \label{scons_time} 
 \end{eqnarray}
 Using the equations in (\ref{scons_time}) and (\ref{phi2_HT}) one can in principle solve $\Lambda_1, \Lambda_3, \Lambda_4$. But it turns out that they are not independent and so $\Lambda_1$ remains undetermined. The solution is
 \begin{eqnarray}
 \nonumber &&
 \Lambda _3 = \frac{\Lambda _1 A}{N(\tau )}-\frac{\Lambda  a^2 N(\tau )^2+A^2+N(\tau )^2}{2 a}, \nl
 \Lambda _4 = -\frac{6 \Lambda _1 a A}{N(\tau )^2}.
 \end{eqnarray} \\ 
 Thus the constraint chain stops here. Existence of the undetermined multiplier $\Lambda_1$ signals that there is gauge symmetry present in the system which we know as the diffeomorphism. The primary first class constraint can emerge out if we take the field redefinition ($N = \sqrt{\dot{t}^2 - \dot{a}^2}$) as worked out in \cite{BMP_rt}. But here we only need the second class constraint to go through the process as first class constraints also can be made second class by incorporating the gauge conditions.  
 Now we solve the constraint $\Phi_2$ and $\Phi_4$ to  remove the unphysical variables $\lambda_a$ and its corresponding momenta $\Pi_{\lambda_a}$. 
 We solve the second class constraint $\Phi_2$ using $\Psi_2$ form (\ref{eh_scons2}) and get
   \begin{equation}
   \Pi_a= 0.
   \end{equation}
   As the second class constraints are directly related to the degrees of freedom count we should always remove them in pairs. So we choose also $\Psi_1$ for removal. The canonical Hamiltonian after solving $\Psi_2$ is
  \begin{equation}
  H_{ghost-free}= -\frac{3 a A^2}{N(\tau )} -a N(\tau) \left(\Lambda  a^2+3\right) . \label{eh_gf}
  \end{equation}
  From the above  equation (\ref{eh_gf}),  we clearly can see that  the Hamiltonian is free from the unwanted momenta and hence do not have any negative norm states. 
  
  The expression for the Hamiltonian is interesting as it matches with $\Psi_1$ (\ref{eh_scons1}). This is expected as it is the very well known Hamiltonian constraint.  One should remember that  at this stage we have to consider the Dirac brackets to replace all calculations involving Poission brackets. Since all the constraints have become second class, they are just identities and hence one can be replaced with respect to the others, to have a viable representation of the theory. To compute the Dirac brackets between the variables, we below list  all the second class constraints (after removing the Lagrange multiplier fields $(\lambda_a , \Pi_{\lambda_a})$).
	 
	 \begin{eqnarray}
	 \nonumber &&
	 S_1 = \frac{3 a^2 A}{N(\tau)^2}+\Pi _N \nl
	 S_2 -= \Pi _A-\frac{3 a^2}{N(\tau)} \nl 
	 S_3 = -\Lambda  a(\tau )^2-\frac{3 A(\tau )^2}{N(\tau)^2}-3 \nl 
	 S_4 = \Pi _a(\tau ) \nl 
	 \end{eqnarray}
Degrees of freedom count is necessary for the fact to validate the theory. In this theory the degrees of freedom in the reduced phase space is 
2 $\times$ (total number of phase space variables) - (2 $\times$ number of first class constraints + number second class constraints) = 2 $\times$ 4 - (2 $\times$ 0 + 6 ) =2. This degrees of freedom count agrees with the standard gravitational results. The matrix $\Delta_{ij} = \{S_i , S_j\}$ 
is given by
\begin{center}
 \[
\begin{bmatrix}
0            &     0  & 0      & \frac{6 a A}{N(\tau )^2}\\
0             &    0  & -\Pi _a & -\frac{6 a}{N(\tau )} \\
0             & \Pi _a & 0 &     -2 \Lambda  a^2 N(\tau ) \\
-\frac{6 a A}{N(\tau )^2} & \frac{6 a}{N(\tau )} & 2 \Lambda  a^2 N(\tau ) & 0
\end{bmatrix}
\]
\end{center}
  Below we list all the nonzero Dirac brackets (\ref{DB_def}) between the fields

\begin{eqnarray}
\nonumber &&
\left\{N(\tau),\Pi _N \right\}_D = 1 \nl 
\left\{a,\Pi _N \right\}_D = -\frac{1}{N(\tau )^5} \left( 3 a^2 (2 a (2 A^2 (\Lambda  N(\tau )^4+3 )+3 N(\tau )^2)+A N(\tau )^3 \Pi _a )\right)  \nl 
\left\{a,\Pi _a \right\}_D= \frac{1}{N(\tau )^4}\left(  6 a A N(\tau )^3 \Pi _a+36 a^2 (A^2+N(\tau )^2)+4 \Lambda ^2 a^4 N(\tau )^6+N(\tau )^4\right)  \nl 
\left\{a,\Pi _A\right\}_D = 2 a^2 \left(3 a A \left(2 \Lambda +\frac{3}{N(\tau )^4}\right)-\Lambda  N(\tau ) \Pi _a \right) \nl 
\left\{A,\Pi _N \right\}_D = \frac{6 a A^2 \Pi _a }{N(\tau )^2}\nl 
\left\{A,\Pi _a \right\}_D = -2 \Lambda  a^2 N(\tau ) \Pi _a \nl 
\left\{A, \Pi _A \right\}_D = -\frac{6 a A \Pi _a}{N(\tau )}+\Pi _a^2+1 \nl 
\{N(\tau ),a\}_D=\frac{6 a A}{N(\tau )^2} \nl 
\{ a,A \}_D= A \Pi _a +\frac{6 a}{N(\tau )} \nl 
\left\{\Pi _{N},\Pi _{a}\right\}_D=\frac{6 \Pi _a \left(\Lambda  a^4 N(\tau )^2-6 a^2 A^2\right)}{N(\tau )^3} \nl 
\left\{\Pi _{N},\Pi _{A}\right\}_D=\frac{3 a^2 \Pi _a \left(6 a A -N(\tau ) \Pi _a \right)}{N(\tau )^3} \nl 
\left\{\Pi _{A},\Pi _{a}\right\}_D=\frac{6 a \Pi _a \left(6 a A-N(\tau ) \Pi _a \right)}{N(\tau )^2}
\end{eqnarray}
The Dirac brackets obtained thus can be used for quantisation of the system in the reduced phasespace.
 \section{Singularity free Gauss Bonnet gravity}
In this section we consider the a matter field added to the Einstein action along with a Gauss-Bonnet term. The action is 
 \begin{eqnarray}
 \nonumber 
 S &=& \int   \sqrt{-g} \left( \frac{R}{2} +\frac{1}{2}\partial_\mu{\phi}\partial^\mu{\phi} - \alpha \xi(\phi) (R_{\mu\nu\rho\sigma}R^{\mu\nu\rho\sigma} - 4R_{\mu\nu}R^{\mu\nu} + R^2)\right)  d^4x.   \label{gb_action}
\end{eqnarray}
In the language of the metric components for the minisuperspace universe (\ref{metric}) the required Lagrangian is  
\begin{eqnarray}
\nonumber 
L &=&\frac{r^2 \sin \theta }{2 \sqrt{1-K r^2} N(\tau )^4} \Big( 6 a (K N(\tau )^5-N(\tau )^3 a'^2 )+a^3 N(\tau )^5 \phi '^2 \\ && + 48 \alpha  \xi (\phi) (K N(\tau )^2-a'^2) (N(\tau ) a''-a' N'(\tau )) + 6 a^2 N(\tau )^2 (a' N'(\tau )-N(\tau ) a'' ) \Big) \label{gb_lag}
\end{eqnarray}

To get a singularity free model of (\ref{gb_lag}) we take $\xi(\phi) = \phi^2$ and $\alpha = 1/32$ as in \cite{kanti} for a negatively curved universe (K=-1). Immediately, we can write down the  the first order Lagrangian with the redefinition (\ref{redefinition})    
\begin{eqnarray}
\nonumber &&  
\mathcal{L} = \frac{1}{N(\tau)^4} \Big(-12 a^2 N(\tau)^2 \left(N(\tau) A'-A N'(\tau )\right)-12 a N(\tau)^3 \left(A^2+N(\tau)^2\right)+2 a^3 N(\tau)^5 \phi '^2 \\ &&
 -3 \phi ^2 \left(A^2+N(\tau)^2\right) \left(N(\tau) A'-A N'(\tau )\right)\Big) +\lambda _a \left(A-a'\right). 
\label{gb_lag_redef}
\end{eqnarray}
Here this $\lambda_a$ is the Lagrange multiplier corresponding to compensate the redefinition of the variables in (\ref{redefinition}). While deriving the form of the Lagrangian (\ref{gb_lag_redef}) we did not consider terms proportional to $( r, \theta)$ as they can be integrate out which eventually will not be effective in the equation of motion.  The Euler-Lagrange equations for this lagrangian are given by 
\begin{eqnarray}
 \nonumber &&
 a(\tau):\frac{1}{N(\tau)^2} \Big(N(\tau) \left(-24 a A'+L \lambda _a'+6 N(\tau)^2 \left(a^2 \phi '^2-2\right)\right)+24 a A N'(\tau )-12 A^2 N(\tau)\Big) = 0,
 \nl  
  A(\tau):\frac{1}{N(\tau)^3} \Big(-24 a N(\tau)^2 \left(A-a'\right)+N(\tau)^3 \lambda _a+6 A^2 \phi  \phi '+6 N(\tau)^2 \phi  \phi '\Big)=0 ,
   \nl
   N(\tau):\frac{1}{N(\tau)^4}\Big( {2 \left(-6 a N(\tau)^2 \left(2 A a'-A^2+N(\tau)^2\right)+a^3 N(\tau)^4 \phi '^2-3 A \phi \left(A^2+N(\tau)^2\right) \phi '\right)} \Big) =0 ,\nl
   \phi:\frac{1}{N(\tau)^4} \Big(6 \phi  \left(A^2+N(\tau)^2\right) \left(A N'(\tau )-N(\tau) A'\right)\Big) - 12 a^2 N(\tau) a' \phi '-4 a^3 N'(\tau ) \phi '-4 a^3 N(\tau)\phi'' =0,
   \nl 
    \lambda_a(\tau) : A - a' = 0.
\end{eqnarray}
\subsection{Hamiltonian formulation and ghost free Hamiltonian}
 In the phasespace,  momenta defined by the equations (\ref{pri1}, \ref{pri2}) are given by
\begin{eqnarray}
&& \nonumber
\Pi _{L}=\frac{1}{N(\tau)^4}\Big(12 a^2 A N(\tau)^2+3 A \phi^2 \left(A^2+N(\tau)^2\right)\Big),
 \nl
 \Pi _{a}=-\lambda _a,
 \nl 
 \Pi _{A}=\frac{1}{N(\tau)^4}\Big(-12 a^2 N(\tau)^3-3 N(\tau) \phi^2 \left(A^2+N(\tau)^2\right)\Big) ,
 \nl
 \Pi _{\phi}=4 a^3 N(\tau) \phi ',
 \nl
 \Pi _{\lambda _a}=0.
\end{eqnarray}  
These momenta give the following   primary constraints which are  
\begin{eqnarray}
&& 
\Phi _1=\Pi _N(\tau)-\frac{1}{N(\tau)^4} \Big(12 a^2 A N(\tau)^2+3 A \phi^2 \left(A^2+N(\tau)^2\right)\approx 0, \label{gb_pcons1}
\\ && 
\Phi _2=\lambda _a + \Pi _a \approx 0, \label{gb_pcons2}
\\ && 
\Phi _3=\Pi _A+ \frac{1}{N(\tau)^4} \Big(12 a^2 N(\tau)^3+3 N(\tau) \phi^2 \left(A^2+N(\tau)^2\right)\approx 0. \label{gb_pricons3}
\\ && 
\Phi _4=\Pi _{\lambda _a}\approx 0, \label{gb_pcons4}
\end{eqnarray}
  We list here the nonzero PBs between the primary constraints
 \begin{eqnarray}
 && \nonumber
 \{\Phi_1,\Phi_2\}=-\frac{24 a A}{N(\tau)^2}, \nl
 \{\Phi_2,\Phi_3\}=-\frac{24 a}{N(\tau)}, \nl
 \{\Phi_2,\Phi_4\}=1.
 \end{eqnarray} 
 A careful redefinition of $\Phi_1 \rightarrow \Phi_1' = \Phi _1-\frac{ 24 a A}{N(\tau)^2}\Phi _4 $ gives more compact PBs which are (only nonzero components are shown)
   \begin{eqnarray}
   && \nonumber
    \{\Phi_2,\Phi_3\}=-\frac{24 a}{N(\tau)}, \nl
   \{\Phi_2,\Phi_4\}=1.
   \end{eqnarray} 
 We can immediately write down the canonical Hamiltonian as  
 \begin{eqnarray}
 H_{can} = \frac{12 a \left(A(\tau )^2+N(\tau)^2\right)}{N(\tau)}+A \Pi _a+\frac{\Pi _{\phi }^2}{8 a^3 N(\tau)}. \label{gb_canham}
 \end{eqnarray}
 The canonical Hamiltonian (\ref{gb_canham}) contains a term linear in momenta $\Pi_a$. This signals the presence of Ostrgradski ghost.  This is in conformity with the general equation (\ref{canham1}) which says corresponding to each higher derivative field there will be at least one term linear in momenta. The total hamiltonian is
 \begin{equation}
 H_{tot}=H_{can}+\Lambda _1 \Phi'_1+\Lambda _2 \Phi_2+\Lambda _3 \Phi_3 + \Lambda _4 \Phi_4.
 \end{equation}
 Equating the PBs of the total hamiltonian with the $\Phi_4$ to zero we get $\Lambda_2 =0$. Also the PBs of the total Hamiltonian with  $\Phi'_1$ and $\Phi_3$ identically give secondary constraints respectively as:
 \begin{eqnarray}
 && 
 \Psi _1=96 a^4 N(\tau)^3 \left(A^2+N(\tau)^2\right)+\Pi _{\phi } \left(12 A \phi \left(A^2+N(\tau)^2\right)-N(\tau)^3 \Pi _{\phi }\right), \label{gb_scons1}
  \\ &&
  \Psi _2=2 a^3 N(\tau)^4 \Pi _a-3 \phi  \Pi _{\phi } \left(A^2+N(\tau)^2\right). \label{gb_scons2} 
 \end{eqnarray}
 On the other hand   PBs of the total Hamiltonian with  $\Phi_2$  in (\ref{gb_pcons2}) gives the following equation
 \begin{equation}
 \frac{3}{8 N(\tau)} \left(32 \left(2 \Lambda _3 a + A^2+N(\tau)^2\right)-\frac{\Pi _{\phi }{}^2}{a^4}\right)-\Lambda _4 =0 \label{scons2}
 \end{equation}
 To get the tertiary constraints it is necessary to see the PBs of the total Hamiltonian with the secondary constraints. Poission bracket of $\Psi_1$ and $\Psi_2$ with the total Hamilonian give the following equations
 \begin{eqnarray}
 \nonumber &&
 -384 a^6 N(\tau)^7 \left(A+\Lambda _2\right) \left(A^2+N(\tau)^2\right)- 96 a^7 N(\tau)^6 (\Lambda _1 \left(3 A^2+5 N(\tau)^2\right) \nl
 + 2 \Lambda _3 A N(\tau))- 3 A N(\tau)^3 \left(A^2+N(\tau)^2\right) \Pi _{\phi }^2+ 3a^3 (-24 A \phi^2 \left(A^2+N(\tau)^2\right)^2 \left(\Lambda _1 A-\Lambda _3 N(\tau)\right)-
 \nl 
4 N(\tau)^3 \phi  \Pi _{\phi } \left(\Lambda _1 A \left(N(\tau)^2-A^2\right)+2 \Lambda _3 N(\tau) \left(2 A^2+N(\tau)^2\right)\right)+\Lambda _1 N(\tau)^6 \Pi _{\phi }^2 ) =0
\nl
96 a^6 N(\tau)^6 \left(-2 \Lambda _1 a A+2 \Lambda _3 a N(\tau)+A^2 N(\tau)+N(\tau)^3\right)+ 
\nl 
72 a^3 \phi^2 \left(A^2+N(\tau)^2\right)^2 \left(\Lambda _1 A-\Lambda _3 N(\tau)\right)+ 
8 a^5 N(\tau)^7 \lambda _a \left(4 \Lambda _1 a+3 N(\tau) \left(A+\Lambda _2\right)\right)+ 
\\ && 
24 a^3 N(\tau)^4 \phi  \Pi _{\phi } \left(\Lambda _3 A+\Lambda _1 N(\tau)\right)+ 
3 N(\tau)^3 \Pi _{\phi }^2 \left(-a^2 N(\tau)^4+A^2+N(\tau)^2\right) =0. \label{tot_secon}
 \end{eqnarray}
 From these two equations of (\ref{tot_secon}) along with (\ref{scons2}) we can easily solve $\Lambda_1$, $\Lambda_2$ and $\Lambda_3$. So, the constraint chain ends here.
 
 As we got the full constraint structure, we notice that the ghost creating momenta appears in (\ref{gb_scons2}). Following the prescription described in the earlier section we  solve the secondary constraints  $\Psi_1$ and $\Psi_2$ to remove the variables $\phi$ and $\Pi_a$ respectively. Thus ghost free canonical Hamiltonian for (\ref{gb_action}) is given by the simple form as 
 \begin{equation}
 H_{ghost-free} = \frac{12 a \left(A(\tau )^2+N(\tau)^2\right)}{N(\tau)}+A \Pi _a+\frac{\Pi _{\phi }^2}{8 a^3 N(\tau)}.
 \end{equation}
 The momenta $\Pi_\phi$ appearing here is of quadratic power and hence is bounded from below. The above result is very interesting which gives back  the ghost free  canonical Hamiltonian  that does  not contain any term linear in any of the  momenta. As we have solved the second class constraints (\ref{scons2}), we need to give up the Poission brackets. Notice, to obtain the ghost free version which was our sole aim is achieved  just by solving $\Psi_1$ and $\Psi_2$. 
 
 The dirac bracket structure can be obtained by solving the constraints $\Phi_1, \Phi_2, \Psi_1, \Psi_2$. Reduction of $\Phi_1$ and $\Phi_2$ are trivial as they do not modify the Poission brackets. Also, by solving these two constraints we  get rid of the unphysical degrees of freedoms  $\lambda_a, \Pi_{\lambda_a}$.  To reduce the phase space further we obtain the Poission brackets between $\Psi_1$ and $\Psi_2$ which is given by 
 \begin{equation}
 \{\Psi_1 , \Psi_2 \} = 6 N(\tau)^3 \left(A^2+N(\tau)^2\right) \left(128 a^6 N(\tau)^4-\Pi _{\phi }^2\right) \label{pb_psi}
 \end{equation}
 To compute the Dirac brackets (\ref{DB_def}) between two canonical functions we shall use (\ref{pb_psi}).   
 Below we give all the non zero Dirac brackets between the variables in the phase space-
 
 \begin{eqnarray}
 \nonumber 
 \left\{N,\Pi _N\right\}_D &=& 1 
 \\ \nonumber 
 \left\{a,\Pi _N\right\}_D &=&-\frac{a^3 N(\tau)^2 \left(32 a^4 \left(3 A^2 N(\tau)+5 N(\tau)^3\right)+\Pi _{\phi } \left(8 A \phi -N(\tau) \Pi _{\phi }\right)\right)}{\left(A^2+N(\tau)^2\right) \left(128 a^6 N(\tau)^4-\Pi _{\phi }^2\right)}
  \\ \nonumber 
 \left\{a,\Pi _a\right\}_D &=& \frac{\Pi _{\phi }^2}{\Pi _{\phi }^2-128 a^6 N(\tau)^4}
 \\ \nonumber 
 \left\{a,\Pi _A\right\}_D &=& -\frac{4 a^3 N(\tau) \left(16 a^4 A N(\tau)^3+\phi  \left(3 A^2+N(\tau)^2\right) \Pi _{\phi }\right)}{\left(A^2+N(\tau)^2\right) \left(128 a^6 N(\tau)^4-\Pi _{\phi }^2\right)}
 \\ \nonumber  
 \left\{a,\Pi _{\phi }\right\}_D &=& -\frac{4 a^3 A N(\tau) \Pi _{\phi }}{128 a^6 N(\tau)^4-\Pi _{\phi }^2}
 \\ \nonumber 
\left\{\phi ,\Pi _N(\tau)\right\}_D  
&=&
 \frac{1}{6 N(\tau) \left(A^2+N(\tau)^2\right) \left(128 a^6 N(\tau)^4-\Pi _{\phi }^2\right)}  
  \left(288 a^4 \phi  (8 A^2 N(\tau)^2 
 +3 A^4+5 N(\tau)^4) 
 \right.\nl \left.
 +  16 a^3 N(\tau) \Pi _a   
   (6 A N(\tau)^2 \phi  
  +6 A^3 \phi-
     N(\tau)^3 \Pi _{\phi })  + 3 \phi (N(\tau)^2-3 A^2) \Pi _{\phi }^2\right) 
\\ \nonumber 
\left\{\phi ,\Pi _a\right\}_D &=& \frac{2 a^2 \left(N(\tau) \Pi _a \left(6 A N(\tau)^2 \phi +6 A^3 \phi -N(\tau)^3 \Pi _{\phi }\right)+96 a \phi  \left(A^2+N(\tau)^2\right)^2\right)}{\left(A^2+N(\tau)^2\right) \left(128 a^6 N(\tau)^4-\Pi _{\phi }^2\right)}
\\ \nonumber 
\left\{\phi ,\Pi _A\right\}_D &=& 
\frac{2 \phi \left(48 a^4 A N(\tau)^3 \left(A^2+N(\tau)^2\right)+\Pi _{\phi } \left(A N(\tau)^3 \Pi _{\phi }+6 A^2 N(\tau)^2 \phi+3 A^4 \phi +3 N(\tau)^4 \phi\right)\right)}{N(\tau)^3 \left(A^2+N(\tau)^2\right) \left(128 a^6 N(\tau)^4-\Pi _{\phi }^2\right)}
\\ \nonumber 
\left\{\phi ,\Pi _{\phi }\right\}_D &=& \frac{128 a^6 N(\tau)^4}{128 a^6 N(\tau)^4-\Pi _{\phi }^2}
 \end{eqnarray}
One should keep in mind that now we are working in the reduced phase space and due to the two second class constraint reduction the dimensionality of the phase space has also reduced by two.

 \section{Conclusion}
 Quantisation of the gravitational fields are one of the most important challenge for this era of theoretical physicists. Problem is that the gravity theories are not renormalisable. Although, adding higher derivative fields can make them renormalisable \cite{stelle} but not all combination of higher derivative terms are allowed \cite{muneyuki}. With these restrictions, higher derivative gravity theories are still considered as strong candidate for developing quantum gravity.  The problem comes along frequently while quantising the higher derivative theories is that there appears negative norm states which we refer to as ghost states \cite{woodard1}. The origin of these ghost fields can be traced back in the canonical Hamiltonian where the momenta corresponding to ghost field appear linearly\cite{majumdar}. For the higher derivative theories only, the Ostrogradsky's theorem itself tells us that the degenerate theories contains ghosts while the non-degerate theories are secretly stable. This issue of removing the ghosts that appearing the caonical momenta is considered here and we showed how one can remove the ghost creating momenta one by one tactfully by considering the constraints only.

	 In this paper, we take simultaneously two model: one consist of only the gravity theory i.e. the Einstein Hilbert action whereas in another model  we considered the Gauss-Bonnet gravity along with a matter field coupled to the Einstein Hilbert action\cite{rizos,kanti}. {Reason for inclusion of  matter field is to confirm that the algorithm provided here do not breaks down even in presence of matter fields.} Following the Hamiltonian formulation we found out all the constraints in the theory. The canonical Hamiltonian, as usual, contains the linear momenta that sources the instability. We notice that these momenta also appear in the second class constraints. So to remove them from the canonical Hamiltonian we solved the second class constraints and found out the canonical Hamiltonian that became independent of the any ghost creating momenta. Thus following the very effective method of Dirac's constraint analysis we construct the ghost free canonical Hamiltonian \cite{dirac}. Further, we compute the Dirac brackets between the canonical fields by solving the second class constraints that contains the ghost crating momenta. It is to be mentioned that one always need to solve  even number of second class constraints to compute Dirac brackets. The degrees of freedom count was done from the number of constraints and this agrees with the expected results.  
	 
	  Owing to this one might inquire  about the system with first class constraints. The first class constraints can be made second class by incorporating gauge conditions. {In fact, in \cite{BMP_rt} we, with other co-authors, discussed the Einstein Hilbert action where there exist primary first class constraint and  gauge condition was proposed to remove it.  The Hamiltonian obtained thereby was free from the linear momenta. }  As future projects, the method followed in this paper can be utilized while quantising theories with more complicated actions. In this regard, one should first check if the canonical Hamiltonian contains ghost creating momentum or not and if it is the constraints can be removed as described in this paper. This is so because by construction the momenta corresponding to the higher derivative fields will appear in some of the primary constraints and thus appear in the canonical Hamiltonian.   


\begin{thebibliography}{999}
	
	\bibitem{stelle} K.~S. ~Stelle, \textit{``Renormalization of higher-derivative quantum gravity"} Phys. Rev. D {\bfseries 16} (1977) 953.
	
	\bibitem{muneyuki} K. Muneyuki, N. Ohta, \textit{``Unitarity versus Renormalizability of Higher Derivative Gravity in 3D"}, Phys. Rev. D  85 (2012) 101501(R).
	
	\bibitem{buch} I.L. Buchbinder, S.D. Odintsov, I.L. Shapiro, \textit{`` Effective action in quantum gravity"}1992, Published in Bristol, UK: IOP (1992) 413 p.
		
	\bibitem{pert} F. Moura, R. Schiappa \textit{``Higher-Derivative Corrected Black Holes: Perturbative Stability and Absorption Cross-Section in Heterotic String Theory"} Class.Quant.Grav. \textbf{24} (2007) 361.
	
	\bibitem{nesterenko} V.V.Nesterenko, ``\textit{Singular Lagrangians with higher derivatives}"J. Phys. A {\bf{ 22}} (1989) 1673.
	
						 R. D. Pisarski, ``\textit{Field theory of paths with a curvature-dependent term}"Phys. Rev. D {\bf{34}} (1986) 670.
						 
						 M.S. Plyushchay, ``\textit{Massive relativistic particle with rigidity} "Int.J.Mod.Phys. A\textbf{4} (1989) 3851;  Nuclear Physics B ``\textit{ The model of the relativistic particle with torsion}" \textbf{362} (1991) 54.
	\bibitem{string} Kristin Forger, Burt A. Ovrut, Stefan J. Theisen, Daniel Waldram \textit{``Higher-Derivative Gravity in String Theory"} Phys.Lett. B \textbf{388} (1996) 512.
		
	J. T. Liu, R. Minasian	\textit{``Higher-derivative couplings in string theory: dualities and the B-field"} Nucl. Phys. B \textbf{874} (2013) 413.
	
	L.N. Granda, S.D. Odintsov \textit{``Effective average action and nonperturbative renormalization group equation in higher derivative quantum gravity "} Grav.Cosmol. \textbf{4} (1998) 85.
	\bibitem{myers} R. C. \textit{``Myers Higher-derivative gravity, surface terms, and string theory"}  
	Phys. Rev. D \textbf{36} (1987) 392. 
	
	\bibitem{codello} A. Codello and R. Percacci,  \textit{``Fixed Points of Higher-Derivative Gravity" } 
	Phys. Rev. Lett. \textbf{97} (2006) 221301.
	
	\bibitem{mignemi} S. Mignemi and D. L. Wiltshire \textit{`` Black holes in higher-derivative gravity theories"} Phys. Rev. D \textbf{46} (1992) 1475.
	 
	\bibitem{camps} J. Camps \textit{``Generalized entropy and higher derivative gravity"} JHEP \textbf{03} (2014) 070.
	
	\bibitem{ostro} M. Ostrogradsky, \textit{Mem. Ac. St. Petersbourg} {\bf V 14} (1850) 385.
	
	\bibitem{woodard1}R.~P.~Woodard,\textit{``Avoiding Dark Energy with $1/R$ Modifications of Gravity"} 	Lect.\ Notes Phys.\  {\bf 720}  (2007) 403.
	
	\bibitem{maldacena} J. Maldacena, \textit{``Einstein Gravity from Conformal Gravity"}, arXiv:1105.5632.
	
	\bibitem{bender} C. M. Bender, S. F. Brandt, J.-H. Chen, and Q. Wang, \textit{``Ghost busting: PT-symmetric interpretation of the Lee model"}Phys. Rev. D 71 (2005) 025014 . 
	
	Carl M. Bender and Philip D. Mannheim \textit{``No-Ghost Theorem for the Fourth-Order Derivative Pais-Uhlenbeck Oscillator Model"} Phys. Rev. Lett. \textbf{100} (2008) 110402.
	
	\bibitem{chen} T. Chen, M. Fasiello, E. A. Lim, A. J. Tolley \textit{``Higher derivative theories with constraints: exorcising Ostrogradski's ghost"} JCAP \textbf{02}(2013)042.
	
	\bibitem{hassan} S. F. Hassan, Rachel A. Rosen,``\textit{Confirmation of the Secondary Constraint and Absence of Ghost in Massive Gravity and Bimetric Gravity}"  JHEP 1204 (2012) 123.
	
	\bibitem{massive_gf}
		
		S. Foffa, M. Maggiore, E. Mitsou, \textit{`` Apparent ghosts and spurious degrees of freedom in non-local theories"} Phys.Lett. B\textbf{733} (2014) 76.
		
		N. A. Ondo, A. J. Tolley \textit{``Complete Decoupling Limit of Ghost-free Massive Gravity"} JHEP \textbf{1311} (2013) 059.
		
		K.  Hinterbichler, JHEP \textit{``Ghost-Free Derivative Interactions for a Massive Graviton"} \textbf{1310} (2013) 102; C. Deffayet, J. Mourad, G. Zahariade, JCAP \textit{``Covariant constraints in ghost free massive gravity"} \textbf{1301} (2013) 032.
	
	\bibitem{volkov} M. S. Volkov \textit{``Stability of Minkowski space in ghost-free massive gravity theory"} Phys.Rev. D \textbf{90} (2014)  024028.
	
	\bibitem{majumdar} L. Buoninfantea \textit{``Ghost and singularity free theories of gravity"} Master's Thesis, arXiv:1610.08744. 
		
	Tirthabir Biswas, Aindriú Conroy, Alexey S Koshelev and Anupam Mazumdar \textit{``Generalized ghost-free quadratic curvature gravity"}  Class. Quantum Grav. \textbf{31}(2014) 015022.
	
	\bibitem{cho} Y.M. Cho, Ishwaree P. Neupane, P.S. Wesson \textit{`` No ghost state of Gauss-Bonnet interaction in warped backgrounds"} Nucl. Phy. B 621 (2002) 388.
	
	\bibitem{langlois} D. Langlois and K. Noui \textit{``Degenerate higher derivative theories beyond Horndeski: evading the Ostrogradski instability"} JCAP02(2016)034.
	\bibitem{amendola}  L. Amendola, C. Charmousis and S. C Davis  \textit{``Constraints on Gauss Bonnet gravity in dark energy cosmologies"} JCAP \textbf{12} (2006) 020.
		
		S. Nojiri, S. D. Odinsov, M. Sasaki , \textit{``Gauss-Bonnet dark energy"} Phys. Rev. D \textbf{71} (2005)123509.
	   S. Nojiri, S. D. Odintsov,  \textit{``Modified Gauss-Bonnet theory as gravitational alternative for dark energy"} Phys.Lett. B \textbf{631} (2005) 1.
	\bibitem{lidsey} J. E. Lidsey and N. J. Nunes \textit{``Inflation in Gauss-Bonnet brane cosmology"}  Phys. Rev. D \textbf{67}(2003) 103510.
	
	
	
	 \bibitem{maeda} H. Maeda \textit{``Gauss-Bonnet braneworld redux: A novel scenario for the bouncing universe"} Phys. Rev. D \textbf{85} (2012) 124012.
	 
	\bibitem{hendi}  S.H. Hendi, S. Panahiyan and B. Eslam Panah \textit{``Charged black hole solutions in Gauss-Bonnet-massive gravity"} JHEP \textbf{01} (2016) 129.
	\bibitem{lidsey1} J. E. Lidsey, S. Nojiri, S. D. Odintsov \textit{``Braneworld Cosmology in (Anti)--de Sitter Einstein--Gauss--Bonnet--Maxwell Gravity"} JHEP 0206:026,2002.
	\bibitem{GB} J. B. Jim\'enez and Tomi S Koivisto \textit{``Extended Gauss-Bonnet gravities in Weyl geometry"} Class. Quantum Grav. \textbf{31 } (2014)135002.
		
		Martin B. Einhorn and D. R. Timothy Jones \textit{``Gauss-Bonnet coupling constant in classically scale-invariant gravity"}
		Phys. Rev. D 91 (2015) 084039. 
		
		S. H. Hendi, S. Panahiyan, B. E. Panah, M. Faizal, and M. Momennia
		\textit{``Critical behavior of charged black holes in Gauss-Bonnet gravity's rainbow"} Phys. Rev. D 94(2016) 024028. 
		
		Claudia de Rham and Samuel Webster \textit{``High-energy theory for close Randall-Sundrum branes"}
		Phys. Rev. D 71(2005) 124025. 
		
		\bibitem{nupane} I. P. Nuepane, \textit{``Gauss-Bonnet assisted braneworld inflation in light of BICEP2 and Planck data"}	Phys. Rev. D \textbf{90} (2014) 123534 .
		
		
		
		\bibitem{charmousis} C. Charmousis and A. Padilla \textit{``The instability of vacua in Gauss-Bonnet gravity"} JHEP \textbf{12} (2008) 038. 
		
			S. Deser and B. Tekin \textit{``Energy in generic higher curvature gravity theories" }
			Phys. Rev. D \textbf{67}(2003) 084009. 


	\bibitem{rizos} J. Rizos, K. Tamvakis, \textit{``On the existence of singularity free solutions in quadratic gravity"} Phys.Lett. B \textbf{326} (1994) 57.
	
	\bibitem{kanti}  	P. Kanti, J. Rizos, K. Tamvakis, \textit{``Singularity-free cosmological solutions in quadratic gravity}" Phys. Rev. D \textbf{59}(1999) 083512 .
	
	\bibitem{ADM} R. Arnowitt, S. Deser, C. W. Misner \textit{``Republication of: The dynamics of general relativity"} Gen Relativ Gravit (2008) 40: 1997. 
	
	\bibitem{cordero1} R. Cordero, A. Molgado, E. Rojas, \textit{``Ostrogradski approach for the Regge-Teitelboim type cosmology"} Phys. Rev. D \textbf{79} (2009) 024024.
	
	\bibitem{BMP_rt} R. Banerjee, P. Mukherjee, and B. Paul \textit{``New Hamiltonian analysis of Regge-Teitelboim minisuperspace cosmology"}
	Phys. Rev. D \textbf{89} (2014) 043508.
	 
	\bibitem{BMP} R. ~Banerjee, P. ~Mukherjee, B. ~Paul,``\textit{Gauge symmetry and W-algebra in higher derivative systems}" JHEP \textbf{1108} (2011) 085  [arXiv : 1012.2969].
		
	\bibitem{deriglazov} Alexei Deriglazov \textit{``Classical Mechanics-Lagrangian and Hamiltonian formamlsm"} 2nd edition, Springer International Publishing Switzerland 2017. 
	\bibitem{dirac} P.A.M. Dirac, Can. J. Math. {\bf 2} (1950) 129;  {\it Lectures on Quantum Mechanics}, Yeshiva University, 1964.
	A.~Hanson, T.~Regge, C.~Tietelboim, ``\textit {Constrained Hamiltonian System}'', (Accademia Nazionale Dei Lincei, Roma, 1976).
    H. J. Rothe, K. D. Rothe, ``\textit{Classical And Quantum Dynamics of Constrained Hamiltonian Systems}'' , World Scientific  Lecture Notes in Physics - Vol. 81.
	\bibitem {sunder} K.~Sundermeyer, ``\textit {Lecture Notes in Physics 169, Constrained Dynamics}'', (Springer-Verlag, 1982).
	 M. ~Henneaux, C. ~Teitelboim, ``\textit{ Quantization of Gauge Systems}'', Princeton University Press.
  D. M.~Gitman and I. V.~Tyutin, `` \textit{Quantization of Fields with Constraints}'',
	{\it  Springer -- Verlag, Berlin, Heidelberg (1990) 291 p}.
	

\end{thebibliography}
\end{document}